\documentclass{article}

\usepackage{amssymb,amsfonts,amsmath,stmaryrd}
\usepackage{cite,enumerate,float,indentfirst}
\usepackage{color}

\def\be{\begin{eqnarray}}
\def\ee{\end{eqnarray}}
\def\nn{\nonumber}

\def\p{\partial}

\def\Tr{{\rm Tr}\,}

\def\OP{\hbox{OP}}
\def\SP{\hbox{SP}}
\def\QQ{Q\text{ Schur\ }}
\def\THL{\widetilde{\text{HL}}}

\definecolor{red}{rgb}{1,0,0}
\definecolor{orange}{rgb}{1,0.5,0}
\definecolor{violet}{rgb}{0.7,0,1}

\newcommand{\Aut}{\mathrm{Aut}}
\newcommand{\Arf}{\mathrm{Arf}}

\hoffset= - 1.0in         

\begin{document}

\title{\vspace{-.5cm}{\Large {\bf  Cut-and-join structure and integrability for spin Hurwitz numbers}}
\author{
{\bf A.Mironov$^{a,b,c}$}\footnote{mironov@lpi.ru; mironov@itep.ru},
\ {\bf A.Morozov$^{d,b,c}$}\thanks{morozov@itep.ru}\ \ and
{\bf S.Natanzon$^{e,b}$}\footnote{natanzons@mail.ru}
}
\date{ }
}

\maketitle

\vspace{-5cm}

\begin{center}
\hfill FIAN/TD-03/19\\
\hfill IITP/TH-07/19\\
\hfill ITEP/TH-11/19\\
\hfill MIPT/TH-06/19
\end{center}

\vspace{2cm}

\begin{center}
$^a$ {\small {\it Lebedev Physics Institute, Moscow 119991, Russia}}\\
$^b$ {\small {\it ITEP, Moscow 117218, Russia}}\\
$^c$ {\small {\it Institute for Information Transmission Problems, Moscow 127994, Russia}}\\
$^d$ {\small {\it MIPT, Dolgoprudny, 141701, Russia}}\\
$^e$ {\small {\it HSE University, Moscow, Russia}}
\end{center}

\vspace{.5cm}

\begin{abstract}
Spin Hurwitz numbers are related to characters of the Sergeev group,
which are the expansion coefficients of the $Q$ Schur functions, depending on odd times
and on a subset of all Young diagrams.
These characters involve two dual subsets: the odd partitions (OP)
and the strict partitions (SP).
The $Q$ Schur functions $Q_R$ with $R\in \SP$ are common eigenfunctions
of cut-and-join operators $W_\Delta$ with $\Delta\in \OP$.
The eigenvalues of these operators are the generalized Sergeev characters, their algebra is isomorphic to the algebra of $Q$ Schur functions.
Similarly to the case of the ordinary Hurwitz numbers, the generating function
of spin Hurwitz numbers is a $\tau$-function of an integrable
hierarchy, that is, of the BKP type. At last, we discuss relations of the Sergeev characters with matrix models.
\end{abstract}

\section{Introduction}

This month it is exactly ten years from the publication of \cite{MMN}
which introduced the commutative ring of general cut-and-join operators
with linear group characters as common eigenfunctions and
symmetric group characters as the corresponding eigenvalues.
Since then, these operators have found a lot of applications in mathematical
physics, from matrix models to knot theory, and led to a crucially important
and still difficult notion of Hurwitz $\tau$-functions.
A variety of further generalizations was considered, from $q,t$-deformations \cite{qtMM}
to the Ooguri-Vafa partition functions \cite{LMOV} and various non-commutative extensions \cite{AMMN,IMM2}. One of the most important generalizations is a construction of open Hurwitz numbers \cite{MMNbg}:
an infinite-dimensional counterpart of the Hurwitz theory realization of algebraic open-closed string model \'a la Moore and Lizaroiu equipped with the Cardy-Frobenius algebra, the closed and open sectors being represented by conjugation classes of permutations and the pairs of permutations, i.e. by the algebra of Young diagrams and bipartite graphs respectively.

Note that the original construction essentially involves the characters of linear groups and symmetric groups (another manifestation of the Schur-Weyl duality) understood as embedded into the linear group $GL(\infty)$ and the symmetric group $S_\infty$.
However, an obvious direction of changing this group set-up remained poorly explored.
In the present paper, we discuss this interesting subject with the hope that
it would add essential new colors to the picture and give rise to many new
applications. That is, instead of the Schur polynomials (characters of linear groups) we deal with the $Q$ Schur functions, and instead of the symmetric groups we deal with the Sergeev groups. Immediate subjects to address within this context are now more or less standard, we list them in the table below indicating where they are discussed in this paper:

\begin{center}
\begin{tabular}{|c|c|c|}
\hline
Subject&ordinary case&spin case\\
\hline
Hurwitz numbers&s.3.1&s.4.1\\
Related symmetric functions&s.3.2.1&s.4.2\\
Frobenius formula&s.3.2.2&s.4.3\\
Algebra of cut-and-join operators $W_\Delta$&s.3.3&s.5\\
Isomorphism of algebra of $W_\Delta$ to (shifted) symmetric functions&s.3.4&s.6\\
Integrability&s.3.5&s.7\\
Matrix models and character expansions&s.3.6&s.8\\
\hline
\end{tabular}
\end{center}

\section{\QQ polynomials}

\subsection{Definitions}

The central role in this paper will be played by somewhat mysterious
\QQ polynomials $Q_R\{p\}$, which depend only on {\it odd} time-variables
$p_{2k+1}$ and only on {\it strict} Young diagrams
$R=\{r_1>r_2>\ldots >r_{l_R}>0\}\in \SP$
(for ordinary diagrams some lines can have equal lengths, i.e.
there is $\geq$ rather than $>$).
These polynomials have two complementary origins:
\begin{itemize}
\item[{\bf (A)}] They were introduced by I. Schur \cite{Schur} in the study of {\it projective}
representations of symmetric groups \\
\item[{\bf (B)}] They were identified by I. Macdonald \cite{Mac} with the Hall-Littlewood polynomials $\hbox{HL}_R$
at $t^2=-1$:
\be
\tilde
Q_R\{p\} =2^{l_R}\cdot
{\cal M}_R\{q=0,t^2=-1,p\} \equiv
2^{l_R}\cdot\THL_R\{p\}
\label{defQ}
\ee
where ${\cal M}_R$ is the Macdonald polynomial (the tilde in $\THL$ denotes restriction to $t^2=-1$,
while the tilde over $Q$ refers to the normalization factor, which will be changed
in the main part of the paper, see (\ref{normal}) at the end of this section). Hereafter, we replace the parameters in the Macdonald book \cite{Mac} $(q,t)\to (q^2,t^2)$.
\item[{\bf (C)}] Their coefficients are expressed through the characters of the
Sergeev group \cite{Serg,Sergrev}.
\end{itemize}
The formal definition of the \QQ polynomials can be found in s.4.2.

\subsection{Immediate corollaries}

Definition (A) implies various determinant (actually, Pfaffian) formulas,
definition (B) implies connection to representation theory, in particular,
the ring structure:
\be
{\bf (B1)}: \ \
{\cal M}_{R_1}\{p\}\cdot {\cal M}_{R_2}\{p\} =
\!\!\sum_{R\in R_1\otimes R_2}\!\! {\cal N}_{R_1,R_2}^R(q,t)\cdot {\cal M}_R
\ \ \ \ \Longrightarrow \ \ \
Q_{R_1}\{p\}\cdot Q_{R_2}\{p\} = \!\!\!\sum_{{R \in R_1\otimes R_2}\atop{R\in {\tiny \SP}}}\!\!\!
N_{R_1,R_2}^R Q_R\{p\}
\ee
A peculiar property of symmetric polynomials from Macdonald family is that the sum at the r.h.s
is restricted from the naive $R_1+ R_2\leq R\leq R_1\cup R_2$
in the lexicographical ordering
to a narrower sum of irreducible representations of $SL_N$ emerging in the tensor product of representations associated with the Young diagrams $R_1$ and $R_2$: $R\in R_1\otimes R_2$
(for example, $[2]\otimes [1,1]$ does {\it not} contain $[2,1]$,
see \cite{MMkerov} for definitions and details).
Macdonald's observations were that
\begin{itemize}
\item[{\bf (B2)}] $\THL_R\{p\}$ for $R\in \SP$ depend only on odd time-variables $p_{2k+1}$
\item[{\bf (B3)}] $\THL_R\{p\}$ for $R\in \SP$ form a sub-ring, i.e.
${\cal N}_{R_1,R_2}^R$ vanish for $R\notin \SP$, provided $q=0,t=i$ and $R_1,R_2\in \SP$. \\
\end{itemize}

\noindent
Note that $\THL_R\{p\}$ do not vanish for $R\notin \SP$, and then they can also
depend on even $p_{2k}$, thus the set of $Q_R\{p\}$ is not the same as the {\it set}
of $\THL_R$, it is a {\it sub-set}, and a {\it sub-ring}.

\bigskip

One more important observation is that \\

\noindent
{\bf (B4)}
after a peculiar rescaling of the Macdonald scalar product \cite{Mac},
\be
\Big<p_\Delta\Big|p_{\Delta'}\Big> = \frac{z_\Delta \delta_{\Delta,\Delta'}}{2^{l_\Delta}}
\label{scapr}
\ee
the restricted HL polynomials for $R\in \SP$ acquire a very simple norm:
\be
\Big<\THL_R\Big|\THL_{R'}\Big> = ||\THL_R ||^2\cdot \delta_{R,R'} = 2^{-l_R}\cdot\delta_{R,R'} \ \ \ \ \ \ {\rm for} \ \ \ \ R,R'\in \SP
\label{orthoHLI}
\ee
\\

\noindent
Actually relevant for the \QQ polynomials is the restriction to odd times,
i.e. the Young diagram $\Delta$ in (\ref{scapr}), which defines the monomial
$p_\Delta = \prod_{i}^{l_\Delta} p_{\Delta_i}$
should have all the lines of odd length:   $\Delta\in \OP$.
Therefore of crucial importance is the celebrated one-to-one correspondence
between  \fbox{the sets of  $\SP$ and $\OP$}.
For example, coinciding are the generating functions
\be
\Sigma_{{\footnotesize \SP}}(q) = \prod_n (1+q^n) = \prod_n(1-q^{2n+1})^{-1} = \Sigma_{{\footnotesize \OP}}(q)
=  \ \ \ \ \ \ \ \ \\
 = 1+ q+q^2+2q^3+2q^4+3q^5+4q^6+5q^7+6q^8+8q^9+10q^{10}+\ldots
\nn
\ee
(this is Sylvester theorem, which is a well known supersymmetric identity).

\subsection{Properties: comparative list}

In this paper, we extend the parallelism between $Q$ and Schur-Macdonald calculus much further:
to the modern fields of integrability and cut-and-join $W$-operators.
Surprisingly or not, the next step, towards Virasoro-like constraints and matrix/network models
fails, at least at the naive level.
This happens even if we do not insist on eigenvalue integrals with Vandermonde-like measures,
but use a ``softer" definition of \cite{MM,MMchar}, making partition function $Z$ directly
of characters.
The reason for this is a puzzling at the moment.

A comparison table of properties looks as follows: note that the Schur and Hall-Littlewood
polynomials are two {\it unrelated} subsets in the Macdonald family.
The $Q$ polynomials belong to the second subset, but are the ones that look
most similar to the first one. For reader's convenience, we provide a short list of the first restricted $\THL$ polynomials in the Appendix.

\bigskip

\centerline{
{\footnotesize
$
\begin{array}{|c||c||c|c|c|c|  }
\hline
&&&&& \\
\text{properties}\backslash \text{polynomials}&\text{Schur  } \ S_R &
{\cal M}_R &
\hbox{HL}_R &
\THL_R  &
\QQ \ Q_R   \\
&&&&& \\
\hline\hline
&&&&& \\
\text{characters} & + & - & - & - & + \\
&&&&& \\
\hline
&&&&& \\
\text{simple determinant formulas} & + & - & - & - & +\\
&&&&& \text{ (Pfaffian)} \\
&&&&& \\
\hline
&&&&& \\
\text{basis in linear space of} \ p & + & + & + & + & +\\
&&&&& \text{ (odd times)}\\
&&&&& \\
\hline
&&&&& \\
\text{closed ring}: R \in R_1\otimes R_2 & + & + & + & + & + \\
&&&&& \\
\hline
&&&&& \\
\text{eigenfunctions of} \ W & + & + & + & + & + \\
\text{(algebra of cut-and-join operators)}
&\text{(differential)} & \text{(difference)} &  &  & \text{(differential)} \\
&&&&& \\
\hline
&&&&& \\
\text{integrability} & + & - & - & - & + \\
& \text{(KP)} & & & & \text{(BKP)} \\
&&&&& \\
\hline
&&&&& \\
\text{dilatation constraint on $Z$ (anomaly)} &
+ & + & + & + & - \\
&&&&& \\
\hline
&&&&& \\
\text{full set of Virasoro-like constraints on } \ Z
& + & + & + & + & - \\
&&&&& \\
\hline
&&&&& \\
\text{eigenvalue integrals} & + & + & + & + & - \\
&& \text{(Jackson)} &&& \\
&&&&& \\
\hline
\end{array}
$
} }

\bigskip

An additional mystery comes from the apparent relevance of the
shift
\be
{\rm Shift}: \ \ \
r_i-i \longrightarrow r_i
\ee
in many formulas for
Schur polynomials: it {\it helps} to convert them into formulas for $Q$.
However, it is not {\it just} this substitution,
some other things should also be adjusted, their is no a
universal conversion rule.
In fact, the shift a sort of converts the ordinary Young diagrams into the strict ones,
but again not quite: the image is not always a Young diagram.
Still, when it is, the shifted diagram belongs to $\SP$.

The difficulties with matrix model formulation seem related to the old problem
of finding a matrix model with only odd time-variables.
Originally it was related to the matrix model solutions of KdV (rather than KP) hierarchy,
and a possible solution was provided by the Kontsevich model,
at the price of making an {\it a priori} non-obvious "Fourier/Miwa transform"
from time-variables to ``the external field".
We are still lacking a clear understanding of this procedure,
which remains a piece of art,
and problems with the \QQ polynomials seem to be a manifestation of this lacuna
in our knowledge.
There are numerous claims that the BKP hierarchy, in variance with the KdV one,
is easier to describe by matrix models, but we did not manage to find
a $Q$-based matrix model on this way.

\subsection{Hamiltonians}

As the Macdonald polynomials, $\THL_R$ are eigenfunctions of the
Calogero-Ruijsenaars-like Hamiltonian\footnote{See a discussion of this and higher Hamiltonains in \cite{MMgenM} and references therein.} ($S_R\{p_k\}$ denotes here the Schur polynomial, which is a symmetric function of variables $x_i$, as a function of power sums $p_k=\sum_i x_i^k$)
\be
\hat{ H} =
\oint\frac{dz}{z}\exp\left(\sum_{k=1} \frac{(1-t^{-2k})z^kp_k}{k}\right)
\exp\left(\sum_{k=1}\frac{q^{2k}-1}{z^k}\frac{\p}{\p p_k}\right)
= \nn \\
= \sum_{m=0} t^{-2m}\cdot S_{[m]}\Big\{(t^{2k}-1)p_k\Big\}
\cdot S_{[m]}\left\{(q^{2k}-1)k\frac{\p}{\p p_k}\right\}
\nn \\
\hat H {\cal M}_R\{p\} = \lambda_R {\cal M}_R\{p\}
\ \ \ \ {\rm with} \ \ \ \
\frac{\lambda_R -1}{t^2-1} = \sum_{i=1}^{l_R} \frac{q^{2r_i}-1}{t^{2i}}
\label{HamCR}
\ee
which, for $q=0$ and $t^2=-1$, reduces to
\be
\hat {\cal H} = \sum_{m=0}  (-)^m\cdot S_{[m]}\Big\{2p_{2k-1}\Big\}
\cdot S_{[1^m]}\left\{ k\frac{\p}{\p p_k}\right\}
= 1 -2p_1\frac{\p}{\p p_1} + p_1^2 \left(\frac{\p^2}{\p p_1^2} - 2\frac{\p}{\p p_2}\right)
+ \ldots
\label{CaRuHTL}
\ee
where the first Schur polynomial depends on odd times only, while the second one
involves derivatives w.r.t. {\it all} times.
All eigenvalues trivialize to
\be
\lambda_R=(-)^{l_R}
\ \ \ \ \ {\rm i.e.} \ \ \ \hat{\cal H} (\THL_R) = (-)^{l_R}\cdot \THL_R
\ee
The Hamiltonian (\ref{HamCR}) is actually a difference operator, since it involves shifts of $p$-variables,
but the Macdonald polynomials are also eigenfunctions of {\it differential}
$W$-operators, which, however, look more involved \cite{Mormac}.

The operators (\ref{CaRuHTL}) are nicely acting on $Q_R$, which depend only on odd times.
However, there is a conspiracy allowing them to act properly also on the
other $\THL_R$, with $R\notin \SP$.

\bigskip

In fact, the Hamiltonian (\ref{HamCR}) becomes the Ruijsenaars one in terms of the Miwa variables, $p_k=\sum_i^nx_i^k$. Moreover, one can write down a set of $n$ integrable Ruijsenaars Hamiltonians in these variables as difference operators acting on the functions of $n$ variables $x_i$ as
\be\label{Hamk}
\hat H_kF(x_i)=\sum_{i_1<\ldots<i_k}{\prod_{m=1}^kD(t^2,x_{i_m})\Delta(x)\over \Delta(x)}\prod_{m=1}^kD(q^2,x_{i_m})F(x_i)
\ee
where $\Delta(x)=\prod_{i<j}(x_i-x_j)$ is the Vandermonde determinant, and $D(\xi,x_i)$ is the operator of dilation of the variable $x_i$: $x_i\to\xi x_i$. The Macdonald polynomials ${\cal M}_R$ are eigenfunctions of these Hamiltonians, while the generating function of the eigenvalues
\be
\sum_k\lambda_R^{(k)}z^k=\prod_{i=1}\Big(1+zq^{2R_i}t^{2(n-i)}\Big)
\ee
In particular,
\be
\lambda_R^{(1)}={\lambda_R t^{2n}-1\over t^2-1}
\ee
where $\lambda_R$ is given in (\ref{HamCR}).

Note that the Hamiltonians (\ref{Hamk}) still depend on the parameter $q$ even at the point $t=q$, while the eigenfunctions, which are the Schur polynomials, do not. This allows one to bring $q$ to zero, obtaining from the difference Hamiltonians the differential ones, which are nothing but the Calogero Hamiltonians\footnote{Similarly, in order to obtain the Jack polynomials from the Macdonald ones, one can bring both $t$ and $q$ to zero together, keeping $\beta:=\log t/\log q$ finite. In this case, one still obtains the Calogero Hamiltonians with $\beta$ being the coupling constant.}.

In the Hall-Littlewood case $q=0$, the Hamiltonians reduce to
\be
\hat H_kF(x_i)=\sum_{i_1<\ldots<i_k}{\prod_{m=1}^kD(t^2,x_{i_m})\Delta(x)\over \Delta(x)}\prod_{m=1}^kD(0,x_{i_m})F(x_i)
\ee
which means that the corresponding $x_i$ at the r.h.s are just put zero. The generating function of
the eigenvalues is, in this case,
\be
\sum_k\lambda_R^{(k)}z^k=\prod_{i=l_{_R}-1}^n\Big(1+zt^{2(n-i)}\Big)
\ee
In particular, upon putting $t^2=-1$, one obtains
\be
\lambda^{(1)}_R={1-(-1)^{n-l_{_R}}\over 2},\ \ \ \ \ \ \lambda^{(2k+1)}_R=\lambda^{(1)}_R\cdot \lambda^{(2k)}_R\nn\\
\lambda^{(2k)}_R=(-1)^k\cdot\prod_{j=1}^n{\xi_R-j+2\over j},\ \ \ \ \ \ \ \ \hbox{where } \xi_R:=\Big[{n-l_{_R}\over 2}\Big]
\ee
and $[\ldots ]$ denotes the integer part. As soon as these eigenvalues depend only on the number of lines in the Young diagram $R$, they essentially differ from the cut-and-join operators of s.\ref{caj}.

\subsection{Application to Hurwitz numbers}

This will be the main topic of the text below,
and the final summary will be given as a comparative table in sec.\ref{summary}.
Here we enumerate the main technical statements, which are discussed
in the middle part of the text.
\begin{itemize}
\item[{\bf 1.}]  Interplay between the skew symmetric functions and finite group characters.
\item[{\bf 2.}] An equivalence of the two definitions of the Hurwitz numbers:
through the enumeration of ramified coverings (``a geometric definition")
and through the Frobenius formula via the symmetric or Sergeev group characters and
Schur functions (``an algebraic definition").
\item[{\bf 3.}] An expression for the skew counterpart of $d_{R}$ in (\ref{dR}) through the (shifted) symmetric functions.
\item[{\bf 4.}] A relation of integrability with the theory of symmetric functions.
\item[{\bf 5.}] The theory of cut-and-join operators $W_\Delta$.
\end{itemize}
In sec.\ref{ordHur} we remind all these issues for the ordinary Hurwitz numbers,
and the remaining sections describe their direct counterparts in the spin Hurwitz case.

\subsection*{Notation}

Below in the text we use the normalization
\be
Q_R = q^{-l_R/2}\cdot\tilde Q_R = q^{l_R/2}\cdot\THL_R
\label{normal}
\ee
so that the polynomials $Q_R$ below have unit norm w.r.t. (\ref{scapr}).

\section{Hurwitz numbers and their properties
\label{ordHur}}

\subsection{Geometric set-up\label{cts}}
The Hurwitz number \cite{Hur,Fro} is a weighted number of globally topologically different branched coverings with the same topological behavior in neighborhoods of critical values. We will consider only coverings over sphere $S^2$. Then a branched covering is given by a continuous map $\varphi: P\rightarrow S^2$, where $P$ is a (not obligatory connected) compact surface. There exists only a finite number $|\Aut(\varphi)|$ of homeomorphisms $f: P\rightarrow P$ such that $\varphi f=\varphi$.

At almost every point $s\in S^2$, there are mapped exactly  $d$ points from $P$. The number $d$ is called degree of $\varphi$. The remaining points are called critical values. There exists only a finite number of critical values. Let $x_1,\dots, x_l$ be all points of $P$ that map to a critical value $s\in S^2$. Running round $x_i$ singly is mapped by $\varphi$ to running round $s$ $\delta_i$ times. Moreover, $\delta_1+\dots+\delta_l=d$. The ordered integers $\delta_i$ represent a partition of $d$, which gives rise to the Young diagram $\Delta_s=[\delta_1,\dots, \delta_l]$ of degree $d$. The diagram $\Delta_s$ is called to be of a topological type $s$.

Consider now the set $V(\Delta_1,\dots,\Delta_k)$ of all branched coverings with critical values $s_1,\dots,s_k\in S^2$ of topological types $\Delta_1,\dots,\Delta_k$. We call the coverings $\varphi^1: P^1\rightarrow S^2$ and $\varphi^2: P^2\rightarrow S^2$ essentially different if there is no a homeomorphism $f: P^1\rightarrow P^2$ such that $\varphi^2 f= \varphi^1$. The Hurwitz number is defined to be \cite{Hur}
\be
\hbox{Hur}_d(\Delta_1,\dots,\Delta_k)=\sum\limits_\varphi\frac{1} {|\Aut(\varphi)|}
\ee
the sum being taken over a maximal set of essentially different coverings from $V(\Delta_1,\dots,\Delta_k)$ with all $|\Delta_i|=d$. It is possible to prove that this number depends only on the Young diagrams $\Delta_1,\dots,\Delta_k$.

The classical Frobenius formula gives a combinatorial expression for the Hurwitz numbers \cite{Hur,Fro},
\be\label{Fro1}
\hbox{Hur}_d(\Delta_1,\dots,\Delta_k)=\frac{[\Delta_1]\dots[\Delta_k]}{(d!)^2} \sum\limits_{R}\frac{\psi_R(\Delta_1)\dots\psi_R(\Delta_k)}
{\psi_R(1)^{(k-2)}}
\ee
where $[\Delta]$ is the number of permutations of the cyclic type $\Delta$,  i.e. the number of elements in the conjugacy class of the symmetric group $\mathfrak{S}_d$ given by the Young diagram $\Delta$, $|\Delta|=d$; $\psi_R(\Delta)$ is value of the character $\psi_R$ of the representation $R$ of the symmetric group $\mathfrak{S}_d$ on the permutation of cyclic type $\Delta$, $\psi_R(1)$ is the value on the permutation with all unit cycles, $\Delta=[\underbrace{1,\ldots,1}_{d\ \rm{times}}]$,
and the sum is taken over all characters of irreducible representations of $\mathfrak{S}_d$.

A definition of more general Hurwitz numbers can be found in \cite{AN}.

\subsection{Algebraic set-up}

\subsubsection{Schur functions and their properties}

The main tool to deal with the Hurwitz numbers and their generating functions is the symmetric functions, that is, the Schur polynomials, and the characters of symmetric groups \cite{Fulton,Mac}.

The Schur polynomials are constructed in the following way. First of all, let us define a set of functions $P_{n}$ by the generating function
\be\label{gf}
\sum_{n}P_{n}z^n:=e^{\sum_k {p_{k}\over k}z^k}
\ee
Now we define the Schur symmetric function for any Young diagram $R$ with $l_{_R}$ lines: $R_1\ge R_2\ge\ldots\ge R_{l_{_R}}$ of size $|R|:=\sum_i R_i$ by the formula
\be
S_R := \det_{i,j} P_{R_i-i+j}
\ee
The Schur functions are orthogonal
\be
\Big< S_R\Big| S_{R'} \Big> =\delta_{R,R'}
\ee
with the scalar product
\be
\Big< p_{k} \Big| p_{l}\Big> = k \cdot \delta_{k,l}
\ee
The Schur functions also satisfy the Cauchy formula
\be
\sum_{R} S_R\{p\}S_{R}\{\bar p\}
= \exp\left(\sum_k \frac{p_{k}\bar p_{k}}{k}\right)
\ee

$S_R\{p\}$ form a full basis in the space of polynomials of $p_{k}$ and thus
form a closed ring.
Let us introduce the Littlewood-Richardson coefficients $N_{R_1R_2}^{R_3}$
\be
S_{R_1}\{p\}S_{R_2}\{p\}=\sum_{R_3}N_{R_1R_2}^{R_3}S_{R_3}\{p\}
\ee
Then, the skew Schur functions $S_{R/P}$, defined as
\be
S_{R}\{p+p'\}=\sum_{P}S_{R/P}\{p\}S_P\{p'\}
\ee
are given by
\be\label{sS}
S_{R/P}\{p\}=\sum_{P}N_{PS}^{R}S_S\{p\}
\ee
The following formulas involving the skew functions are also correct:

\begin{itemize}
\item the Cauchy formula
\be
\sum_{R} S_{R/T_1}\{p\}S_{R/T_2}\{\bar p\}
= \exp\left(\sum_k \frac{p_{k}\bar p_{k}}{k}\right)\cdot \sum_{P} S_{T_1/P}\{\bar p\}S_{T_2/P}\{p\}
\ee
\item the expansion formula
\be
S_{R/T}\{p+p'\}=\sum_{P}S_{R/P}\{p\}S_{P/T}\{p'\}
\ee
\end{itemize}

\subsubsection{Frobenius formula}

Now we can discuss a combinatorial formula for the Hurwitz numbers \cite{Hur,Fro}. First of all,
we need the character of symmetric group in the representation $R$, which value on the element from the conjugacy class $\Delta$, $\psi_R(\Delta)$ is the coefficient of the Schur functions \cite{Fulton}
\be\label{Frob}
S_R=\sum_{|\Delta|=n}{\psi_R(\Delta)\over z_\Delta}p_\Delta
\ee
where we denote $n=|R|$, $p_\Delta:=\prod_i p_{\Delta_i}=\prod_k p_k^{m_k}$, i.e. $m_k$ is the number of lines of length $k$. The number of elements in the conjugacy class of $\Delta$ is $|\Delta|!/z_\Delta$, where $z_\Delta:=\prod_k k^{m_k}m_k!$ is the standard symmetric factor of the Young diagram (order of the automorphism) \cite{Fulton}, while the dimension of the representation $R$ of the symmetric group $\mathfrak{S}_n$ is $n!\cdot d_R$, $d_R=S_R(p_k=\delta_{k,1})$. The quantity $d_R$ is manifestly given by the hook formula
\be\label{dR}
d_R={\psi_R(1^{n})\over n!}={\prod_{i<j}(R_i-i-R_j+j)\over n!(l_{_R}+R_i-i)!}=\prod_{\hbox{all boxes of }R}{1\over\hbox{hook length}}
\ee

As any characters, $\psi_R(\Delta)$ satisfy the orthogonality conditions:
\be
\sum_{\Delta}{\psi_R(\Delta)\psi_{R'}(\Delta)\over z_\Delta}=\delta_{RR'}
\ee
\be\label{OPR}
\sum_{R}{\psi_R(\Delta)\psi_{R}(\Delta')\over z_\Delta}=\delta_{\Delta\Delta'}
\ee
Note that the Littlewood-Richardson coefficients are expressed through the characters $\psi_R(\Delta)$ as
\be\label{LR}
N_{PS}^{R}=\sum_{\Delta_1,\Delta_2}{\psi_P(\Delta_1)\psi_S(\Delta_2)\psi_R(\Delta_1+\Delta_2)\over
z_{\Delta_1}z_{\Delta_2}}
\ee
where $\Delta_1+\Delta_2$ denotes the reordered union of all lines of the two diagrams.

Now the Hurwitz numbers are given as follows ($g$ is the genus of the base, $|\Delta_i|=d$) \cite{Hur,Fro}
\be\label{Fro}
\boxed{
\hbox{Hur}_{g,d}(\{\Delta_i\}):= \sum_{R}  d_R^{2-2g} \prod_i \phi_{R,\Delta_i}}
\ee
where following Frobenius we introduce $\phi_{R,\Delta}:=\psi_R(\Delta)/(z_\Delta d_R)$. This formula at $g=0$ agrees with (\ref{Fro1}), since
\be
[\Delta]={d!\over z_\Delta},\ \ \ \ \ \ \ \psi_R([1^d])=d!\cdot d_R
\ee

\subsection{Cut-and-join (W-) operators and Young diagram algebra}

One can naturally associate with Hurwitz numbers a set of commuting differential operators. These operators generalize the cut-and-join operator of \cite{GD}, which is the simplest one in the whole set, and are constructed in the following way \cite{MMN,MMN1}. They are originally invariant differential operators on the matrices $M$ from $GL(\infty)$, so that the time-variables $p_k = \Tr M^k$, and the eigenvalues of the matrices are related with $p_k$ by the Miwa transformation. Then, the generalized cut-and-join operators are
\be
\hat W_{\!_\Delta} := \frac{1}{z_{_\Delta}}:\prod_i \hat D_{\delta_i}:
\label{Wops}
\ee
and
\be
\hat D_k := \Tr (M \partial_{M})^k
\ee
The normal ordering in (\ref{Wops}) implies that all the derivatives $\partial_M$
stand to the right of all $M$.
Since $W_\Delta$ are invariant matrix operators, and we apply them only to invariants, they can be realized as differential operators in $p_k$ \cite{MMN}. In particular, the simplest cut-and-join operator $\hat{W}_{[2]}$, \cite{GD} is
\be\label{W2}
\hat{ W}_{[2]} ={1\over 2} \sum_{a,b} \Big((a+b)p_ap_b\p_{a+b} + abp_{a+b}\p_a\p_b\Big)
\ee
Another example is
\be
\hat{W}_{[3]} =
\frac{1}{3}\sum_{a,b,c\geq 1}^\infty
abcp_{a+b+c} \frac{\p^3}{\p p_a\p p_b\p p_c}
+ \frac{1}{2}\sum_{a+b=c+d} cd\left(1-\delta_{ac}\delta_{bd}\right)
p_ap_b\frac{\p^2}{\p p_c\p p_d} + \\
+ \frac{1}{3} \sum_{a,b,c\geq 1}
(a+b+c)\left(p_ap_bp_c + p_{a+b+c}\right)\frac{\p}{\p p_{a+b+c}}
\label{W3p}
\ee
An essential property of these generalized cut-and-join operators is that they form a commutative family with the common eigenfunctions being the Schur functions:
\be
\hat W_{\!_\Delta} S_R\{p\} = \phi_{R,\Delta}\cdot S_R\{p\}
\label{evW}
\ee
What is important is that one can lift in this formula the restriction $|\Delta|=|R|$, then, one immediately obtains for the diagram $\Delta$ containing $r$ unit cycles: $\Delta=[\tilde\Delta, 1^r]$,
\be\label{cphi}
\phi_{R,\Delta}=\left\{\begin{array}{cl}
0\ \ \ \ \ \ \ \ \ \ \ \ \ \ \ \ \ \ \ \ \ \ \ \ \ \  &|\Delta|>|R|\\
&\\
\displaystyle{(|R|-|\Delta|+r)!\over r!(|R|-|\Delta|)!}\ \phi_{R,\hat\Delta}=\displaystyle{(|R|-|\Delta|+r)!\over r!(|R|-|\Delta|)!}\
\displaystyle{\psi_{R}(\hat\Delta)\over z_{\hat\Delta}d_R}\ \ \ \ \ \ \ \ \ \ \ \ \ \ \ \ \ \ \ \ \ \ \ &|\Delta|\le |R|
\end{array}\right.
\ee
where $\hat\Delta:=[\Delta,1^{|R|-|\Delta|}]$.

Note that the commutative family of the generalized cut-and-join operators gives rise to the associative algebra of Young diagrams:
\be\label{IKa}
\hat W_{\Delta_1}\hat W_{\Delta_2}=\sum_{\Delta}C^\Delta_{\Delta_1\Delta_2}\hat W_{\Delta}
\ee
Note that this algebra was first constructed in \cite{IK} just in terms of $\phi_{R,\Delta}$. Indeed,
using (\ref{evW}), one can immediately translate (\ref{IKa}) into terms of the vectors $\phi_{R,\Delta}$ in the space of representations $R$ of $S_\infty$:
\be
\phi_{R,\Delta_1}\phi_{R,\Delta_2}=\sum_{\Delta}C^\Delta_{\Delta_1\Delta_2}\phi_{R,\Delta}
\ee
 Still, the fact that the structure constants $C^\Delta_{\Delta_1\Delta_2}$ are independent of $R$ follows in the simplest way from the algebra of commuting cut-and-join operators.

\subsection{$\phi_{R,\Delta}$ and shifted Schur functions}

In accordance with formula (\ref{cphi}), $\phi_{R,\Delta}$ expresses through $\psi_R([\Delta,1^{|R|-|\Delta|}])$. In its turn, the latter can be expressed \cite{IK} through the shifted Schur functions \cite{OO}. Indeed, an explicit formula for $\psi_R([\Delta,1^{|R|-|\Delta|}])$
involves the skew Schur functions at the special point $p_k=\delta_{1,k}$:
\be\label{IK}
\psi_R([\Delta,1^{|R|-|\Delta|}])=
(|R|-|\mu|)!\cdot\sum_{\mu\vdash |\Delta|} S_{R/\mu}\{\delta_{1,k}\}\psi_\mu(\Delta)
\ee
This formula follows from the manifest expression for the skew Schur functions, (\ref{sS}) through the
Littlewood-Richardson coefficients and the manifest expression  (\ref{LR}) for these latter.
Then, using the expansion (\ref{Frob}) for the Schur function
and repeating several times the orthogonality relation of the symmetric group characters (\ref{OPR}),
one immediately obtains (\ref{IK}).

The quantity $S_{R/\mu}\{\delta_{1,k}\}$ can be expressed through the shifted Schur functions $S^*_\mu(R)$ \cite{OO}. The shifted Schur functions are symmetric functions of the $n$ variables $x_i-i$ and can be defined either through the sum over the reverse semi-stable Young tableaux $T$, which entries strictly decrease down the column and non-strictly decrease right in the row,
\be
S^*_\mu(x_i):=\sum_T\prod_{(i,j)\in T}\left(x_{T(i,j)}+i-j\right)
\ee
or through the determinant
\be\label{S*det}
S^*_\mu(x_i)={\det_{i,j}(x_i+n-i;\mu_j+n-j)\over\det_{i,j}(x_i+n-i;n-j)}
\ee
where $(x;n):=\prod_{k=0}^{n-1}(x-k)=x!/(x-n)!$. In the limit of large $x_i$, $(x;n)\to x^n$, and formula (\ref{S*det}) reduces to the formula for the standard Schur polynomials
\be
S_\mu(x_i)={\det_{i,j}x_i^{\mu_j+n-j}\over\det_{i,j}x_i^{n-j}}
\ee
Hence, the standard Schur polynomials are the large $x_i$-asymptotics of the shifted ones.

Equivalently, the shifted Schur functions can be also unambiguously expressed through the shifted power sums
\be
p^*_k:=\sum_i \left[(x_i-i)^k-(-i)^k\right]
\ee
if one requires
\be
S^*_\mu\{p^*_k\}=S_\mu\{p^*\}+\sum_{\lambda:\ |\lambda|<|\mu|} c_{\mu\lambda}S_\lambda\{p^*_k\}\nn\\
\nn\\
S^*_\mu(R_i)=0\ \ \ \ \ \ \ \ \hbox{if }\mu\notin R
\ee
Now one can use \cite[formula (0.14)]{OO}\footnote{In the paper \cite{OO}, the shifted Schur function $S^*_\mu(R)$ is related with the number of skew standard Young tableaux, $d_{R/\mu}=(|R|-|\mu|)!\cdot S_{R/\mu}\{\delta_{1,k}\}$.}
\be\label{SIK}
S^*_\mu(R_i)={S_{R/\mu}\{\delta_{1,k}\}\over d_R}
\ee
in order to obtain finally
\be\label{OO}
\psi_R([\Delta,1^{|R|-|\Delta|}])=(|R|-|\Delta|)!d_R\cdot\sum_{\mu\vdash |\Delta|} S^*_\mu(R_i)\psi_\mu(\Delta)
\ee
and
\be
\phi_{R,\Delta}=\sum_{\mu\vdash |\Delta|} S^*_\mu(R_i)\ {\psi_\mu(\Delta)\over z_\Delta}
\ee

\subsection{Integrability}

Now one can consider the generating function of the Hurwitz numbers $\hbox{Hur}_{g,n}(\Delta_1,\Delta_2,\underbrace{[\Delta,1^{n-m}],[\Delta,1^{n-m}],\ldots,[\Delta,1^{n-m}]}_k)$, with some fixed $|\Delta|=m$:
\be
Z_{g,n}(\beta;p,\bar p):=\sum_{\Delta_1,\Delta_2,k}
\hbox{Hur}_{g,n}(\Delta_1,\Delta_2,\underbrace{[\Delta,1^{n-m}],[\Delta,1^{n-m}],\ldots,[\Delta,1^{n-m}]}_k) p_{\Delta_1}p_{\Delta_2}
{\beta^k\over k!}=\nn\\
=\sum_Rd_R^{-2g}S_R\{p\}S_R\{\bar p\}e^{\beta\phi_{R,[\Delta,1^{n-m}]}}
\ee
One can definitely consider more than two sets of Young diagrams $\Delta_1,\ \Delta_2$ and accordingly more times variables, however, the standard integrability will not persist in those cases. Now, one can further define
\be
Z_g(\beta;p,\bar p):=\sum_n q^nZ_{g,n}(\beta;p,\bar p)
\ee
and we will restrict ourselves only to the genus zero. At last, we use the continuation (\ref{cphi}) of $\phi_{R,\Delta}$ to $|R|\ne|\Delta|$ and consider more than one $\Delta$ in order to obtain finally (see details in \cite{MMN})
\be\label{tau}
Z(\{\beta_i\})=\sum_RS_R\{p\}S_R\{\bar p\}e^{\sum_i\beta_i\phi_{R,\Delta_i}}
\ee
where we have fixed a set of $\{\Delta_i\}$ and rescaled $qp_k\to p_k$.

Now one may ask when the generating function (\ref{tau}) is a $\tau$-function of the KP hierarchy (or, more generally, the Toda hierarchy) w.r.t. to each set of time-variables $p_k$ and $\bar p_k$. First of all, the Schur function satisfies the KP equation:
\be
\forall R \ \ \ \
{\rm and} \ \ \ \ \ u = 2\frac{\p^2}{\p p_1^2} \log \Big(S_R\{p\}\Big)
\nn \\
\frac{\p}{\p p_1} \left(-12\frac{\p u}{\p p_3} + 6u\frac{\p u}{\p p_1}
+ \frac{\p^3 u}{\p p_1^3}\right)
+ 12 \frac{\p^2 u}{\p p_2^2} = 0
\ee
and, in fact, the entire KP hierarchy.
This is nearly obvious from the fermionic realization of characters
\cite{JM,GKLMM}
but in the ordinary formulation looks like a set of non-trivial identities.
Linear combinations
\be
\tau\{p\} =\sum_R c_R\cdot S_R\{p\}
\ee
satisfy the hierarchy, provided the coefficients $c_R$ satisfy quadratic Pl\"ucker relations,
i.e. if $\tau$ satisfies bilinear Hirota equations.
The first KP equation in Hirota form is \cite{Hirota,JM}
\be
\left(D_{[1,1,1,1]}+3D_{[2,2]}-4D_{[3,1]}\right)(\tau\circ\tau)=0
\ee
where
\be
D_\Delta\big(\tau\circ\tau\big):=\left.\prod_{i=1}^{l_{_\Delta}}\ \Delta_i\left({\partial\over\partial p_{\Delta_i}}-{\partial\over\partial
p_{\Delta_i}'}\right)\Big(\tau\{p\}\circ\tau\{p'\}\Big)\right|_{p_k=p_k'}
\ee
while the generating function of the whole hierarchy is written in terms of the generating parameters $y_{k}$ as
\be
\sum_jP_j(-2y)P_j(\tilde D)e^{\sum_i y_{i}D_{i}}\tau\circ\tau=0
\ee
where $P_k$ are the polynomials (\ref{gf}) and
\be
\tilde D_{k}:={D_{k}\over k}
\ee

It was first proved in \cite{GKM2} that the partition function (\ref{tau}) solves the KP hierarchy w.r.t. each set of time-variables $p_k$ and $\bar p_k$ if the sum in the exponential, $\sum_i\beta_i\phi_{R,\Delta_i}$ is an arbitrary linear combination of the Casimir operators, $C_k(R)=\sum_j \Big[(R_j-j)^k-(-j)^k\Big]$. A particular case of this claim \cite{Oko} is the case of only one $\Delta=[2]$, since $\phi_{R,[2]}$ is associated with $C_2(R)$:
\be\label{phi2}
\phi_{R,[1]}=\sum_j R_j=|R|,\ \ \ \ \ \ \phi_{R,[2]}={1\over 2}\sum_j \Big[(R_j-j+1/2)^2-(-j+1/2)^2\Big]
\ee
The $\tau$-functions of this kind are called hypergeometric \cite{OS}.

However, higher $\phi_{R,\Delta}$ are not linear combinations of $C_k(R)$. The proper combinations of $C_k(R)$ are nicknamed the completed cycles. Hence, the final claim is \cite{GKM2}:

\bigskip

\framebox{only the generating function (\ref{tau}) with the completed cycles gives a $\tau$-function of the KP hierarchy}

\bigskip

\noindent
More details and discussion of other cases can be found in \cite{AMMN1,AMMN}.

\subsection{Matrix models and character expansions}

One can rewrite the generating function (\ref{tau}) in the cases, when it is a $\tau$-function, in the form \cite{OS,AMMN}
\be\label{tau2}
Z_w=\sum_RS_R\{p\}S_R\{\bar p\}w_R
\ee
with the function $w_R$ being the product
\be
w_R=\prod_{i,j\in R}f(i-j)
\ee
since exponential of any linear combination of $C_k(R)$ can be presented \cite[sect.3]{AMMN} as $w_R$ with some function $f(x)$. For instance, $e^{\beta C_2(R)}=\prod_{i,j\in R}e^{\beta (i-j)}$.

It turns out that the generating functions (\ref{tau2}) are sometimes partition functions of matrix models. For instance, the partition function of the rectangular $N_1\times N_2$ complex matrix model is \cite{IMM2,MM}
\be\label{Z1}
{\cal Z}_{N_1\times N_2} \{p\} := \sum_{R}
{D_R(N_1)D_R(N_2)\over d_R}\cdot
S_R\{p\}
\ee
where
\be
D_R(N):=S_R\{p_k=N\}
\ee
is dimension of the representation of $SL_N$ group given by the Young diagram $R$.

Similarly, the partition function of the Gaussian Hermitean matrix model is \cite{IMM2,MM}
\be\label{Z2}
Z_N\{p\} := \sum_{R}
\frac{S_R\{p_k=\delta_{k,2}\}D_R(N)}{d_R}\cdot
S_R\{p\}
\ee

Both these partition functions are known to be $\tau$-functions of the KP hierarchy (and the Toda chain hierarchy) \cite{MMMM,IMM2,GMMMO,KMMOZ}, which is evident from the results of the previous subsection: the both partition functions can be presented in the form (\ref{tau2}) with $\bar p_k=N_2$ in (\ref{Z1}) and $\bar p_k=\delta_{2,k}$ in (\ref{Z2}), and the weight function $w_R$ of the form
\be\label{wR}
w_R={D_R(N)\over d_R}=\prod_{i,j\in R}(N+i-j)
\ee

\section{Spin Hurwitz numbers}

\subsection{Geometric set-up}

Spin Hurwitz numbers are similar to the classical Hurwitz numbers adapted to coverings  with spin structures \cite{EOP,G}.

Spin bundle was defined (under the name of theta-characteristic) by B. Riemann as a bundle over Riemann surfaces such that its tensor square is the cotangent bundle \cite{Atiyah,Mum}. The spin bundle on a surface $P$ has an equivalent topological description using a quadratic form (Arf-function) $\omega:H_1(P,\mathbb{Z}_2)\rightarrow\mathbb{Z}_2$ \cite{J,N2,N3}.
Any Arf-function $\omega$ has an algebraic invariant $\Arf(\omega)\in\{0,1\}$. On any oriented topological surface $P$ of genus $h$, there exist $2^{h-1}(2^h+1)$ Arf-functions $\omega$ with $\Arf(\omega)=0$ and $2^{h-1}(2^h-1)$ Arf-functions $\omega$ with $\Arf(\omega)=1$.

Now consider the set $V(\Delta_1,\dots,\Delta_k)$ of all branched coverings
$$\varphi:P\rightarrow S^2$$
with critical values $s_1,\dots,s_k\in S^2$ of topological types $\Delta_1,\dots,\Delta_k$, all $|\Delta_i|=d$ (see sec.\ref{cts}). There is a unique Arf-function on the sphere $S^2$. The covering $\varphi$ transforms this Arf-function into an Arf-function $\omega(\varphi)$ on $P$, if and only if $\varphi$ does not have branch points of even order. Thus $\omega(\varphi)$ exists iff all $\Delta_i\in \OP$.

In this case, the definition of spin Hurwitz number is
\be\label{sHur1}
\hbox{sHur}_d(\Delta_1,\dots,\Delta_k)=\sum\limits_\varphi
\frac{(-1)^{\Arf(\omega(f))}} {|\Aut(\varphi)|}
\ee
where the sum is taken over a maximal set of essentially different coverings from $V(\Delta_1,\dots,\Delta_k)$ with all $|\Delta_i|=d$. It is possible to prove that this number depends only on the Young diagrams $\Delta_1,\dots,\Delta_k$.

It follows from (\ref{sHur1}), \cite{G,Lee} that
\be\label{sHur}
\hbox{sHur}_d(\Delta_1,\dots,\Delta_k)=\frac{(d!)^{k-2}}{z_{\Delta_1}\dots z_{\Delta_k}} \sum\limits_{R\in\hbox{\footnotesize SP}} \frac{\Psi_R(\Delta_1)\dots\Psi_R(\Delta_k)}
{\Psi_R(1)^{(k-2)}}
\ee
where $\Psi_R(\Delta)$ is value of the character $\Psi_R$ of the representation $R$ of the Sergeev group on the permutation of cyclic type $\Delta$, $|\Delta|=R$; $\Psi_R(1)$ is the value on the permutation with all unit cycles, $\Delta=[\underbrace{1,\ldots,1}_{d\ \rm{times}}]$, and the sum is taken over all irreducible representations of the symmetric group $\mathfrak{S}_d$ corresponding to the Young diagrams with pairwise different lengths of lines.

From now on, we treat (\ref{sHur}) as a definition of the spin Hurwitz numbers.

\subsection{Schur $Q$-functions and their properties}

A counterpart of the Schur functions which allows one to construct a combinatorial formula for the spin Hurwitz numbers similar to the Frobenius formula (\ref{Fro}) is the system of symmetric Schur $Q$-functions \cite{Schur,Mac}. These functions were originally introduced by I. Schur on the projective representations of the symmetric groups and turn out to induce characters of the Sergeev group \cite{Serg,Jose,Sergrev}. They can be obtained from the Hall-Littlewood polynomials $\hbox{HL}_R(t)$ \cite{Mac},
\be
Q_R=
\left\{\begin{array}{cl}
2^{l_{_R}/2}\cdot\hbox{HL}_R(t^2=-1)&\hbox{for}\ R\in\hbox{SP}\\
&\\
0&\hbox{otherwise}
\end{array}
\right.
\ee
Hereafter, SP (strict partitions) denotes a set of Young diagrams with all lengths of lines distinct.

However, there is a manifest way to construct them. To this end, let us define a set of functions $Q_{n,m}$ by the generating function
\be\label{gf1}
\sum_{n,m}Q_{n,m}z_1^nz_2^m:=\left(e^{2\sum_k {p_{2k+1}\over 2k+1}(z_1^{2k+1}+z_2^{2k+1})}-1\right){z_1-z_2\over z_1+z_2}
\ee
It is a power series in both $z_1$ and $z_2$, since the exponential in (\ref{gf1}) is equal to 1 at $z_2=-z_1$. Moreover, $Q_{n,m}=-Q_{m,n}$, i.e. the matrix $\mathfrak{Q}_{ij}:=Q_{R_i,R_j}$ associated with a Young diagram $R$ is antisymmetric. The indices of the matrix run from 1 to $l_{_R}$ for even $l_{_R}$ and from 1 to $l_{_R}+1$ for add $l_{_R}$, i.e. we add a line of zero length to the Young diagram with odd number of lines, $Q_{0,n}$ being non-zero.

Now we define the $Q$-Schur symmetric function via the Pfaffian of $\mathfrak{Q}$:
\be
Q_R := 2^{-l_{_R}/2}
\cdot {\rm Pfaff}(\mathfrak{Q})
\ee
With this normalization, for
\be
\Big< p_{2k+1} \Big| p_{2l+1}\Big> = (k+1/2) \cdot \delta_{k,l}
\ee
the $Q$-functions are orthogonal:
\be
\Big< Q_R\Big| Q_{R'} \Big> = ||Q_R||^2\cdot\delta_{R,R'}
\ee
with
\be
||  Q_R||^2 = 1
\ee
The Cauchy formula acquires the form
\be
\sum_{R\in\hbox{\footnotesize SP}} Q_R\{p\}Q_{R}\{\bar p\}
= \exp\left(\sum_k \frac{p_{2k+1}\bar p_{2k+1}}{k+1/2}\right)
\ee

$Q_R\{p\}$ form a full basis in the space of polynomials of $p_{2k+1}$ and thus
form a closed ring.
Since Cauchy formula is true and the norms of $Q$ are unities,
the skew-functions $Q_{R/R'}$ are defined
directly through the structure constants of the ring \cite{MMkerov,Mcauchy}.
Namely, introduce the Littlewood-Richardson coefficients ${\cal N}_{R_1R_2}^{R_3}$ in the standard way
\be
Q_{R_1}\{p\}Q_{R_2}\{p\}=\sum_{R_3\in\hbox{\footnotesize SP}}{\cal N}_{R_1R_2}^{R_3}Q_{R_3}\{p\}
\ee
Then, the skew $Q$-Schur functions $Q_{R/P}$, defined as
\be
Q_{R}\{p+p'\}=\sum_{P\in\hbox{\footnotesize SP}}Q_{R/P}\{p\}Q_P\{p'\}
\ee
are given by
\be
Q_{R/P}\{p\}=\sum_{P\in\hbox{\footnotesize SP}}{\cal N}_{PS}^{R}Q_S\{p\}
\ee
The usual formulas involving the skew functions are also correct:

\noindent
the Cauchy formula
\be
\sum_{R\in {\footnotesize \SP}} Q_{R/S}\{p\}Q_{R/T}\{\bar p\}
= \exp\left(\sum_k \frac{p_{2k+1}\bar p_{2k+1}}{k+1/2}\right)\cdot \sum_{P\in\hbox{\footnotesize SP}} Q_{S/P}\{\bar p\}Q_{T/P}\{p\}
\ee
and the expansion formula
\be
Q_{R/S}\{p+p'\}=\sum_{P\in\hbox{\footnotesize SP}}Q_{R/P}\{p\}Q_{P/S}\{p'\}
\ee

\subsection{Frobenius formula}

Now we are ready to discuss a combinatorial formula for the spin Hurwitz numbers \cite{G,Lee}. First of all,
we associate the characters of the Sergeev group $\Psi_R(\Delta)$ with the coefficients
\be
Q_R=\sum_{\Delta\in \hbox{\footnotesize OP}}{\Psi_R(\Delta)\over z_\Delta}p_\Delta
\ee
where OP (odd partitions) is a set of Young diagrams with all lengths of lines odd\footnote{The both SP and OP have the same dimensions as can be seen from their generation functions: the generation function of number of SP at a given level $n$ is equal to $\prod_n (1+q^n)$, while that of OP is $\prod_n(1-q^{2n+1})^{-1}$, and these two products are equal to each other.}. These coefficients plays for the spin Hurwitz numbers the same role as do the characters of symmetric groups for ordinary Hurwitz numbers. Their particular values are:
\be
\Psi_{[r]}(\Delta)=2^{l_{_\Delta}-1/2},\ \ \ \ \ \ r=|\Delta|\nn\\
\nn\\
\Psi_{R}([1^r])=2|R|!\mathfrak{d}_R,\ \ \ \ \ \ \ r=|R|\nn\\
\nn\\
\Psi_{R}([2k+1])=\left\{\begin{array}{cl}
(-1)^{R_1+1}2^{l_{_\Delta}/2}&\hbox{if }l_{_\Delta}\le 2\\
\\
0&\hbox{if }l_{_\Delta}>2
\end{array}\right.,\ \ \ \ \ 2k+1=|R|
\ee

As any characters, they satisfy the orthogonality conditions:
\be
\sum_{\Delta\in \hbox{\footnotesize OP}}{\Psi_R(\Delta)\Psi_{R'}(\Delta)\over 2^{l_{_\Delta}}z_\Delta}=\delta_{RR'}
\ee
\be
\sum_{R\in \hbox{\footnotesize SP}}{\Psi_R(\Delta)\Psi_{R}(\Delta')\over 2^{l_{_\Delta}}z_\Delta}=\delta_{\Delta\Delta'}
\ee
Note that the Littlewood-Richardson coefficients for the $Q$-functions are expressed through the characters $\Psi_R(\Delta)$ in the usual way
\be
{\cal N}_{PS}^{R}=\sum_{\Delta_1,\Delta_2\in\hbox{\footnotesize OP}}{\Psi_P(\Delta_1)\Psi_S(\Delta_2)\Psi_R(\Delta_1+\Delta_2)\over
2^{l_{_{\Delta_1}}+l_{_{\Delta_2}}}z_{\Delta_1}z_{\Delta_2}}
\ee
where $\Delta_1+\Delta_2$ denotes the reordered union of all lines of the two diagrams.

We will also need the quantity which is a counterpart of the standard $d_R$ and regulates the dimension of representation $R$ of the Sergeev group
\be
\mathfrak{d}_R:={1\over 2}\cdot Q_R\{\delta_{k,1}\}
\ee
It is manifestly given by
\be
\mathfrak{d}_R:=2^{|R|-1-{l_{_R}\over 2}}\left({1\over\prod_j^{l_{_R}}R_j!}\right)\prod_{k<m}{R_k-R_m\over R_k+R_m}
\ee
which is a counterpart of the hook formula (\ref{dR}). It is non-zero only for $R\in \hbox{SP}$.

Now the spin Hurwitz numbers for the genus $g$ base $\Sigma$ with the spin structure $\omega$ such that Arf($\omega)=p$  are given as follows ($|\Delta_i|=d$) \cite{G,Lee}
\be
\boxed{
\hbox{sHur}_{g,n}^{(p)}(\{\Delta_i\}):= 2^{(d-2)(g-1)} \sum_{R\in \hbox{\footnotesize SP}} (-1)^{p\cdot l_{_R}} \mathfrak{d}_R^{2-2g} \prod_i \Phi_R(\Delta_i)}
\ee
where $\Delta\in \hbox{OP}$, $l_R$ is the number of lines in the Young diagram $R$, and $\Phi_R(\Delta):=\Psi_R(\Delta)/(z_\Delta \mathfrak{d}_R)$. As compared with (\ref{Fro1}), this formula contains an additional sign factor and additionally depends on the parity
$p\in\mathbb{Z}/2\mathbb{Z}$ \cite{Atiyah,J,Mum,N2,N3}. The surface $\Sigma=S^2$ has only even spin structure and, therefore, the $\hbox{Hur}_{0,n}^{(1)}$ does not exist. However, all formulas can be smoothly extended also to this case, \cite{Lee}. Then, this formula at $g=0$ agrees with (\ref{Fro1}), since
\be
\Psi_R([1^d])=2\cdot d!\cdot d_R
\ee
In particular,
\be\label{Phi3}
\Phi_R([1^{|R|}])=2\nn\\
\Phi_R([3,1^{|R|-3}])={1\over 6}\sum_i^{l_{_R}}R_i(R_i-1)(R_i-2)-\sum_{i<j}R_iR_j
\ee

\section{Cut-and-join (W-) operators and Young diagram algebra\label{caj}}

$W$-operators are again defined as graded differential operators in time variables $p_k$ with common eigenfunctions being $Q$ Schur functions. They are labeled by $\Delta\in \OP$ so that the order of the operator is $|\Delta|$. One can immediately check that the first operators are
\be
\widehat{\mathfrak{W}}_{[1]} =2\sum_a(2a+1)p_{2a+1}{\p\over\p p_{2a+1}}\nn\\
\widehat {\mathfrak{W}}_{[1,1]} =2\sum_a a(2a+1)p_{2a+1}{\p\over\p p_{2a+1}}+\sum_{a,b}(2a+1)(2b+1)p_{2a+1}p_{2b+1}\frac{\p^2}{\p p_{2a+1}\p p_{2b+1}}
\nn\\
\widehat{\mathfrak{W}}_{[1,1,1]} =
{2\over 3}\sum_{a\ge 1} a(2a+1)(2a-1)p_{2a+1}{\p\over\p p_{2a+1}}
+ \sum_{a,b} (2a+1)(2b+1)(a+b)p_{2a+1}p_{2b+1}\frac{\p^2}{\p p_{2a+1}\p p_{2b+1}}\,+\nn\\
+{1\over 3}\sum_{a,b,c\ge 1}(2a+1)(2b+1)(2c+1)p_{2a+1}p_{2b+1}p_{2c+1}{\p ^3\over\p p_{2a+1}\p p_{2b+1}\p p_{2c+1}}
\ee
and the first non-trivial one is at the third level
\be\label{W3}
2\cdot\widehat{\mathfrak{W}}^{(3)} = \frac{1}{3}\sum_{a,b,c}\left(
 4\cdot (2a+2b+2c+3)(p_{2a+1}p_{2b+1}p_{2c+1}
+ p_{2a+2b+2c+3})\frac{\p}{\p p_{2a+2b+2c+3}}
+ \right.\nn \\ \left.
+  (2a+1)(2b+1)(2c+1)p_{2a+2b+2c+3}\frac{\p^3}{\p p_{2a+1}\p p_{2b+1}\p p_{2c+1}}\right)
+   \sum_{a+b=c+d}  (2c+1)(2d+1) p_{2a+1}p_{2b+1}
\frac{\p^2}{\p p_{2c+1}\p p_{2d+1}}
= \nn \\
=  p_1^2\p_1^2  +\frac{1}{3}\Big(12(p_1^3+p_3)\p_3 + p_3\p_1^3\Big)
+  12 p_3p_1\p_3\p_1 + \Big(20(p_3p_1^2+p_5)\p_5 + 3p_5\p_3\p_1^2\Big)
+ (p_3^2+2p_5p_1)(9\p_3^2 +  10  \p_5\p_1)
+ \ldots
\ee
They are Hermitian with
\be
p_{2k+1}^\dagger = \frac{2k+1}{2}\p_{2k+1}
\ee
The eigenvalues of $\widehat{\mathfrak{W}}_{[1^k]}$ on the eigenfunction $Q_R$ are equal to
\be
\Lambda_R([1])=2|R|\nn\\
\nn\\
\Lambda_R([1,1])=2\cdot{|R|(|R|-1)\over 2}\nn\\
\nn\\
\ldots\nn\\
\nn\\
\Lambda_R([1^k])=2\cdot{|R|!\over (|R|-k)!k!}
\ee
These eigenvalues are nothing but lifting of $\Phi_{R,\Delta}$ to $|R|\ne|\Delta|$ similar to (\ref{cphi}): for the diagram $\Delta$ containing $r$ unit cycles: $\Delta=[\tilde\Delta, 1^r]$,
\be\label{cPhi}
\Phi_{R,\Delta}=\left\{\begin{array}{cl}
0\ \ \ \ \ \ \ \ \ \ \ \ \ \ \ \ \ \ \ \ \ \ \ \ \ \  &|\Delta|>|R|\\
&\\
\displaystyle{(|R|-|\Delta|+r)!\over r!(|R|-|\Delta|)!}\ \Phi_{R,\hat\Delta}=\displaystyle{(|R|-|\Delta|+r)!\over r!(|R|-|\Delta|)!}\
\displaystyle{\Psi_{R}(\hat\Delta)\over z_{\hat\Delta}\mathfrak{d}_R}\ \ \ \ \ \ \ \ \ \ \ \ \ \ \ \ \ \ \ \ \ \ \ &|\Delta|\le |R|
\end{array}\right.
\ee
where $\hat\Delta:=[\Delta,1^{|R|-|\Delta|}]$.

At the same time, the eigenvalue of $\hat {\mathfrak{W}}^{(3)}$ is equal to
\be\label{ev}
\Lambda_R^{(3)}={1\over 6}\sum_i^{l_{_R}}R_i(R_i^2-1)
\ee
This eigenvalue is not equal to $\Phi_R([3,1^{|R|-3}])$ (\ref{Phi3}), as it was in the ordinary Hurwitz case (\ref{cphi}):
\be
\Lambda_R^{(3)}=\Phi_R([3,1^{|R|-3}])+{1\over 2}\Big(|R|-1\Big)|R|
\ee
This means that, in order to construct $\widehat{\mathfrak{W}}_{[3]}$, one has to add $\widehat {\mathfrak{W}}_{[1,1]}$ to $\widehat {\mathfrak{W}}^{(3)}$ so that the eigenvalue of $\widehat {\mathfrak{W}}_{[3]}$ would become exactly $\Phi_R([3,1^{|R|-3}])$:
\be
\widehat{\mathfrak{W}}_{[3]}=\widehat{\mathfrak{W}}^{(3)}+{1\over 2}\widehat{\mathfrak{W}}_{[1,1]}
\ee
Such ``corrected" cut-and-join operator generates the spin Hurwitz numbers, however, the operator $\hat {\mathfrak{W}}^{(3)}$, (\ref{W3}) instead generates the BKP $\tau$-function, see the next section, i.e. provides a counterpart of the completed cycle.

At the first 5 levels, one can obtain the following values of $\Phi_R(\Delta)$:

\bigskip

\noindent
\centerline{\rule{8cm}{.5pt}}
\be
\Phi_R([1])=2|R|\nn
\ee
\centerline{\rule{8cm}{.5pt}}
\be
\Phi_R([1,1])=2\cdot{|R|(|R|-1)\over 2}\nn
\ee
\centerline{\rule{8cm}{.5pt}}
\be
\Phi_R([1,1,1])=2\cdot{|R|(|R|-1)(|R|-2)\over 6}\nn\\
\nn\\
\Phi_R([3])={1\over 6}\sum_i^{l_{_R}}R_i(R_i^2-1)-{1\over 2}\Big(|R|-1\Big)|R|\nn
\ee
\centerline{\rule{8cm}{.5pt}}
\be
\Phi_R([1,1,1,1])=2\cdot{|R|(|R|-1)(|R|-2)(|R|-3)\over 24}\nn\\
\nn\\
\Phi_R([3,1])=(|R|-3)\cdot\Phi_R([3])\nn
\ee
\centerline{\rule{8cm}{.5pt}}
\be
\Phi_R([1,1,1,1,1])=2\cdot{|R|(|R|-1)(|R|-2)(|R|-3)(|R|-4)\over 120}\nn\\
\nn\\
\Phi_R([3,1,1])={(|R|-3)(|R|-4)\over 2}\cdot\Phi_R([3])\nn\\
\nn\\
\Phi_R([5])={1\over 40}\sum_i^{l_{_R}}R_i^5-{6|R|-11\over 24}\sum_i^{l_{_R}}R_i^3-{|R|(5|R|-3)(5|R|-12)\over 60}
\label{HPHI}
\ee
\centerline{\rule{8cm}{.5pt}}

\bigskip

\noindent
Linearly combining $\Phi_R(\Delta)$ with different $\Delta$'s, one can easily cook up the expressions of the form
\be
C_R^{(k)}=\sum_i R_i^k
\ee
which are \boxed{\hbox{\bf counterparts of the completed cycles}} in the ordinary, non-spin case.

\section{$\Phi_{R,\Delta}$ and symmetric functions}

Similarly to (\ref{IK}), one can prove that
\be\label{IKQ}
\boxed{
\Psi_R([\Delta,1^{|R|-|\Delta|}])=(|R|-|\mu|)!\cdot\sum_{\mu\in\hbox{\footnotesize SP}_{|\Delta|}} Q_{R/\mu}(\delta_{1,k})\Psi_\mu(\Delta)}
\ee
where the sum runs over the strict partitions.

Now one could try to express $Q_{R/\mu}(\delta_{1,k})$ through the shifted symmetric functions. Let us start with the shifted Macdonald functions \cite{Ok}, which are symmetric functions of the $n$ variables $x_it^{-2i}$ and can be again defined through the sum over the reverse semi-stable Young tableaux $T$,
\be
M^*_\mu(x_i):=\sum_T \xi_T(q,t)\prod_{(i,j)\in T}t^{2(1-T(i,j))}\left(x_{T(i,j)}-q^{2(j-1)}t^{2(1-i)}\right)
\ee
where $\xi_T(q,t)$ are the same coefficients (rational functions of $q$ and $t$) as in the usual (non-shifted) Macdonald polynomials.

Equivalently, they can be also unambiguously expressed through the shifted power sums
\be
p^*_k:=\sum_i \left[(x_it^{-2i})^k-(t^{-2i})^k\right]
\ee
if one requires
\be\label{Mreq}
M^*_\mu\{p^*_k\}=M_\mu\{p^*\}+\sum_{\lambda:\ |\lambda|<|\mu|} \tilde c_{\mu\lambda}(q,t)M_\lambda\{p^*_k\}\nn\\
\nn\\
M^*_\mu(q^{R_i})=0\ \ \ \ \ \ \ \ \hbox{if }\mu\notin R
\ee
Now one would have to put $q=0$ and $t^2=-1$ in the shifted Macdonald polynomials and consider only the strict partitions in order to obtain the Schur $Q$-function. However, one immediately
realizes that the requirement (\ref{Mreq}) becomes too singular, when one puts $q=0$ and $t^2=-1$, and, besides, the Schur $Q$-functions would become symmetric functions in variables $(-1)^ix_i$.

Instead of this, we consider the usual symmetric functions of variables $x_i$, or functions of variables $p_k=\sum x_i^k$, and define
\be\label{Qreq}
\overline{Q}_\mu\{p_k\}:=Q_\mu\{p_k\}+\sum_{\lambda\in\hbox{\footnotesize SP}:\ |\lambda|<|\mu|} C_{\mu\lambda}Q_\lambda\{p_k\},
\nn\\
\nn\\
\bar Q_\mu(R_i)=0\ \ \ \ \ \ \ \ \hbox{if }\mu\notin R
\ee
Then, as a counterpart of (\ref{SIK}), we obtain
\be\label{QIK}
\overline{Q}_\mu(R_i)=2^{|\mu|}\cdot{Q_{R/\mu}\{\delta_{1,k}\}\over \mathfrak{d}_R}
\ee
in order to obtain finally
\be\label{QOO}
\boxed{
\Psi_R([\Delta,1^{|R|-|\Delta|}])=(|R|-|\Delta|)!\mathfrak{d}_R\cdot\sum_{\mu\in\hbox{\footnotesize SP}_{|\Delta|}}2^{-|\mu|} \overline{Q}_\mu(R_i)\Psi_\mu(\Delta)
}
\ee
This formula can be immediately recast into an explicit expression for $\Phi_{R,\Delta}$:
\be
\boxed{
\Phi_{R,\Delta}=\sum_{\mu\in\hbox{\footnotesize SP}_{|\Delta|}}2^{-|\mu|} \overline{Q}_\mu(R_i)\ {\Psi_\mu(\Delta)\over z_\Delta}
}
\ee
which gives (\ref{HPHI}) in particular examples.

\section{Integrability}

Similarly to the ordinary Hurwitz numbers, one can construct the generating function of the spin Hurwitz numbers. A counterpart of (\ref{tau}) is, in this case,
\be\label{Qtau}
Z^{(p)}(\{\beta_i\};p,\bar p)=\sum_{R\in \hbox{\footnotesize SP}}(-1)^{p\cdot l_{_R}}Q_R\{p\}Q_R\{\bar p\}e^{\sum_i\beta_i\Phi_{R,\Delta_i}}
\ee
One now may ask when it is a $\tau$-function of an integrable hierarchy w.r.t. each set of time-variables $p_k$ and $\bar p_k$.

First of all, $Q$-functions $Q_R\{p\}$ depend only on odd time-variables $p_{2k+1}$,
and one {\it could} think that they have something to do with the KdV equation
and KdV hierarchy.
However,  $Q_R\{p\}$ do not solve KdV.
Instead, they provide solutions to the BKP hierarchy.
The first two BKP equation in the Hirota form are \cite{BKP}
\be
\Big(D_{[1,1,1,1,1,1]}-20D_{[3,1,1,1]}-80D_{[3,3]}+144D_{[5,1]}\Big)(\tau\circ\tau)=0 \nn \\
\Big(D_{[1,1,1,1,1,1,1,1]}+28D_{[3,1,1,1,1,1]}-560D_{[3,3,1,1]}
- 336D_{[5,1,1,1]}-2688D_{[5,3]}+5760D_{[7,1]}\Big)(\tau\circ\tau)=0
\label{BKP1}
\ee
Any particular $Q$-function solves these equations
(and also solves the whole hierarchy \cite{BKPs}).
The generating function of the whole hierarchy is written in terms of the generating parameters $y_{2k+1}$ as
\be
\sum_j{\cal P}_j(2y){\cal P}_j(-2\tilde D)e^{\sum_i y_{2i+1}D_{2i+1}}\tau\circ\tau=0
\ee
where
\be
\exp\Big(\sum_k {p_{2k+1}\over k+1/2}z^k\Big)=\sum_k z^k{\cal P}_k(p),\ \ \ \ \ \ \tilde D_{2k+1}:={D_{2k+1}\over k+1/2}
\ee
This system is associated with the orthogonal Grassmannian.

In complete analogy with the KP case (\ref{tau}), the generating function of the spin Hurwitz numbers is a $\tau$-function of the BKP hierarchy w.r.t. each set of time-variables $p_k$ and $\bar p_k$
not for arbitrary linear combination of $\Phi_{R,\Delta}$ in the exponential of (\ref{Qtau}), but only in the case of arbitrary combination of the Casimir operators. More precisely,  consider the generating function
\be\label{Qtau2}
Z^{p}(\{\beta_i\};p,\bar p)=\sum_{R\in \hbox{\footnotesize SP}}(-1)^{p\cdot l_{_R}}Q_R\{p\}Q_R\{\bar p\}W_R
\ee
Then, one can extract from the Hirota equations (\ref{BKP1}) w.r.t. $p_k$ and $\bar p_k$ the bilinear relations
\be
W_{[1]}W_{[3,2]}=W_{[2,1]}W_{[3]}\\
W_{[3,1]}W_{[2]}=W_{[2,1]}W_{[3]}\\
W_{[2,1]}W_{[3]}=W_{\emptyset}W_{[3,2,1]}\\
\ldots
\ee
These bilinear relations has an evident solution
\be\label{WR}
W_R=\prod_i F(R_i)
\ee
Hence, we arrive at the claim that

\bigskip

\be
\boxed{\hbox{(\ref{Qtau2}) solves the BKP hierarchy if }W_R=\prod_i F(R_i)}
\ee

\bigskip

\noindent
with an arbitrary function $F(x)$. This describes a counterpart of the hypergeometric $\tau$-functions, i.e. $\tau$-functions of the form (\ref{Qtau2}) that satisfy the hierarchy equations w.r.t. to the both sets of time-variables, in the BKP case. Numerous discussions of the BKP hierarchy and related issues can be found in \cite{Orlov}.

Note that, similarly to (\ref{cphi}), one can continue the Sergeev characters to $|R|\ne |\Delta|$. However,
in variance with (\ref{phi2}), the lowest non-trivial Sergeev character $\Phi_R([3])$, (\ref{Phi3}) {\it is not} of the form (\ref{WR}), because of the second mixing term, and, hence, does not give rise to a $\tau$-function. At the same time, the eigenvalue of the first non-trivial cut-and-join operator (\ref{ev}), $\tilde \Phi_R([3])$ {\it is} a linear combination, and can be used in (\ref{Qtau}) in order to obtain a $\tau$-function.

Also note that the formulas for $W_R$, and many similar ones, in the spin case involve the quantities $R_i$, while the same formulas in the non-spin case, $R_i-i$. This is because the shift $R_i-i$ effectively makes the partition $R_i$ strict, and, in the spin case, the partitions are strict from the very beginning.
A particular manifestation of this phenomenon is also seen from the sum over the Young diagrams with restricted numbers of lines
\be\label{Qtau3}
\sum_{R:\ l_{_R}\le N}S_R\{\bar p\}Q_{\hat R}\{p\}W_{\hat R}
\ee
which is a $\tau$-function of the BKP hierarchy w.r.t. time variables $p_k$ and a $\tau$-function of the KP hierarchy w.r.t. time-variables $\bar p_k$. Here $\hat R$ is the strict partition made from $R$: $\hat R_i= R_i-i$.

\section{Matrix models and the character expansions}

Similarly to the KP case, one can study the sums over the $Q$-functions of types (\ref{Z1}) and (\ref{Z2}) in the spin case in attempt to associate them with matrix model partition functions. Hence, we look at the series
\be\label{ZQ2}
Z_N\{p\} := \sum_{R\in {\footnotesize \SP}}
\frac{Q_R\{p_k=\delta_{k,r}\}\cdot Q_R\{p_k=N\}}{Q_R\{p_k=\delta_{k,1}\}}\cdot
Q_R\{p\}
\ee
with some fixed $r$. However, this sum is not a $\tau$-function of the BKP hierarchy at all, which is not surprising, since
\be
\frac{Q_R\{p_k=N\}}{Q_R\{p_k=\delta_{k,1}\}}
\ee
is not a weight of the proper form (as was the case in the Hermitean Gaussian matrix model), (\ref{WR}).
Neither $Q_R\{p_k=N\}$ makes any sense of a representation dimension.

Moreover, one also should not expect
\be\label{ZQ1}
{\cal Z}_{N_1\times N_2} \{p\} := \sum_{R\in {\footnotesize \SP}}
\frac{Q_R\{p_k=N_1\}\cdot Q_R\{p_k=N_2\}}{Q_R\{p_k=\delta_{k,1}\}}\cdot
Q_R\{p\}
\ee
to be a $\tau$-function, because it also has no form (\ref{Qtau3}) with a proper weight. However, it turns that (\ref{ZQ1}) {\it is} a $\tau$-function of the BKP hierarchy (not of the form (\ref{Qtau3}), since it does not survive the deformation $Q_R\{p_k=N\}\to Q_R\{\bar p\}$). One can look for an origin of this sum among various matrix integrals. Instead, one can equally well {\it define} a matrix model by this formula. This is a very clear and simple definition with many standard implications. In particular, one can now wonder if this partition function
\begin{itemize}
\item is $\tau$-function of the BKP integrable  hierarchy (it is as was stated above),
\item satisfies Virasoro-like constraints,
\item admits a $W$-representation,
\item possesses an integral (matrix model) representation.
\end{itemize}
We leave 3 latter issues for future studies.

\section{Conclusion
\label{summary}}

The ordinary Schur functions $S_R\{p\}$ have the following properties:

\begin{itemize}
\item[{\bf a)}] Their coefficients depend on $R$ in  a peculiar way:  they are symmetric functions of $R_i-i$
and form a linear basis in the space of such functions.
\item[{\bf b)}] These coefficients are proportional to eigenvalues of $W$-operators
that form a commutative algebra with non-trivial structure constants.
\item[{\bf c)}] These coefficients are essentially the characters of symmetric group,
depending on the conjugacy class of the group element.
\item[{\bf d)}] The KP $\tau$-function is made of exponentiated Casimir eigenvalues.
\item[{\bf e)}] The two sets of functions do not coincide: the Casimir eigenvalues are associated not with the cyclic classes of
symmetric group elements, but with their linear combinations (one Casimir eigenvalue for each symmetric group).
\end{itemize}

Based on these {\it facts} one can introduce additional (superficial) {\it definitions}:
\begin{itemize}
\item[{\bf f)}] {\it Call} the algebra of $W$-operator an ``algebra of Young diagrams".
Due to a) and b), it is isomorphic to the multiplication algebra of shifted
Schur functions, and, due to c), is associated
with symmetric group characters of $S_\infty$.
\item[{\bf g)}] {\it Call} linear combinations in e) ``completed cycles".
\end{itemize}

\bigskip

This collection of statements and definitions has direct analogues for $Q_R\{p\}$:

$$
\begin{array}{|c|c|c|c|}
\hline
&&&\\
1)&\text{Schur vs characters}
& S_R\{p\}=\sum_\Delta {\psi_R(\Delta)\over z_\Delta}  p_\Delta
& Q_R\{p\} = \sum_{\Delta \in \hbox{\scriptsize OP}} {\Psi_R(\Delta)\over z_\Delta} p_\Delta
\\
&&&\\
2)&\text{integrability}
& \sum_R S_R\{\bar p\}S_R\{p\} e^{\sum_k b_kh_k(R)}
& \sum_{R\in {\scriptsize \SP}} Q_R\{\bar p\}Q_R\{p\} e^{\sum_kB_kH_k(R)}
\\
&&&\\
3)&\text{Hamiltonians (shifted power sums)}
& h_k(R)=\sum_i \big[(R_i-i)^k-(-i)^k\big]
& H_k(R)=\sum_i R_i^k
\\
&&& \\
4)&\text{$W$-eigenfunctions}
& \hat W_\Delta S_R\{p\} = \phi_R(\Delta) S_R\{p\}
& \hat {\mathfrak{W}}_\Delta Q_R\{p\} = \Phi_R(\Delta) Q_R\{p\}
\\
&&&\\
5)&\text{$W$-eigenvalues}
& \phi_R(\Delta) :={\psi_R(\Delta)\over S_R\{\delta_{k,1}\}z_\Delta}\ \hbox{and}\ (\ref{cphi})
& \Phi_R(\Delta) :=2{\Psi_R(\Delta)\over Q_R\{\delta_{k,1}\}z_\Delta}\ \hbox{and}\ (\ref{cPhi})
\\
&&&\\
6)&\text{$W$-eigenvalues and Hamiltonians}
& h_k(R)=\sum_{|\Delta|\le k} a_\Delta\phi_R(\Delta)
& H_k(R)=\sum_{|\Delta|\le k} A_\Delta\Phi_R(\Delta)
\\
&&&\\
7)&\text{``Inverse transformation" }
& \phi_R(\Delta) = f_\Delta(h_k)
& \Phi_R(\Delta) = F_\Delta(H_k)
\\
&&&\\
8)&\text{ Isomorphism with (shifted)}
& \phi_R(\Delta) \in \left\{S^*_\mu(R_i)\right\}
& \Phi_R(\Delta) \in \left\{\overline{Q}_\mu(R_i)\right\}\\
&\hbox{symmetric functions}&&\\
&&&\\
\hline
\end{array}
$$
\paragraph{Comment.} In fact, the Hamiltonians $h_k$ differ from the shifted symmetric power sums in \cite{Ok} by a constant
$(1-2^{-k})\zeta(-k)$.

In the text, we provided detailed explanations of these statements and evidence in favour of them.

\section*{Acknowledgements}

Our work is partly supported by the grant of the Foundation for the Advancement of Theoretical Physics ``BASIS" (A.Mir., A.Mor.), by  RFBR grants 19-01-00680 (A.Mir.) and 19-02-00815 (A.Mor.), by joint grants 19-51-53014-GFEN-a (A.Mir., A.Mor.), 19-51-50008-YaF-a (A.Mir.), 18-51-05015-Arm-a (A.Mir., A.Mor.), 18-51-45010-IND-a (A.Mir., A.Mor.). The publication was partly prepared within the framework of the Academic Fund Program at the National Research University Higher School of Economics (HSE) in 2020-2021 (grant № 20-01-009) and by the Russian Academic Excellence Project "5-100".

\section*{Appendix}

For the reader's convenience we provide  a short list of the first restricted HL polynomials:

\bigskip

\hspace{-.5cm}\centerline{
{\scriptsize
$
\begin{array}{|c|c|c|}
\hline &&\\
\text{level} & \widetilde Q_R & \text{other} \ \THL_R \\
&& \\
\hline
&& \\
1 & \THL_{[1]} = p_{1}=S_{[1]} &\\
&& \\
\hline
&& \\
2 &  \THL_{[2]} = p_{1}^2=S_{[1]}^2 &    \THL_{[1,1]}
= \frac{-p_2+p_1^2}{2}=S_{[1,1]} \\
&& \\
\hline
&& \\
3 & \THL_{[3]} = \frac{ p_3+2p_1^3}{3}, \ \ \ \ \
\THL_{[2,1]} = \frac{-p_3+p_1^3}{3}=S_{[2,1]} &
\THL_{[1,1,1]} = \frac{2p_3-3p_2p_1+p_1^3}{6}=S_{[1,1,1]}
\\
&& \\
\hline
&& \\
4 &  \THL_{[4]} = \frac{ (2p_3+p_1^3)p_1}{3}, \ \ \ \
\THL_{[3,1]} = \frac{(-p_3+p_1^3)p_1}{3}=p_1\,S_{[2,1]}
& \ \ \THL_{[2,2]} = \frac{3p_4-4p_3p_1 +  p_1^4}{6},
\ \ \ \ \   \THL_{[2,1,1]}= \frac{(2p_3-3p_2p_1+p_1^3)p_1}{6} = p_1\,S_{[1,1,1]}
\\
&&\\
&&
\THL_{[1,1,1,1]} = \frac{-6p_4+8p_3p_1+3p_2^2-6p_2p_1^2+p_1^4}{24} =S_{[1,1,1,1]}
\\
&& \\
\hline
\end{array}
$
}}


\newpage


\begin{thebibliography}{12}

\bibitem{MMN} A. Mironov, A. Morozov, S. Natanzon,
Theor.Math.Phys.{\bf 166} (2011) 1-22,  arXiv:0904.4227 \!\!;
    JHEP {\bf 11} (2011) 097,  arXiv:1108.0885 \\
for far-going extensions of cut-and-join structure see \cite{IMM2,IMM}

\bibitem{qtMM} A. Morozov, A. Popolitov, Sh. Shakirov, Phys.Lett. {\bf B784} (2018) 342-344, arXiv:1803.11401\\
R. Lodin, A. Popolitov, Sh. Shakirov, M. Zabzine, arXiv:1810.00761

\bibitem{LMOV} H. Ooguri, C. Vafa, Nucl.Phys. {\bf B577} (2000) 419-438, arXiv:hep-th/9912123\\
K. Liu and P. Peng,
  J. Diff. Geom. {\bf 85} (2010), no. 3 479-525, arXiv:0704.1526;
Math.Res.Lett. {\bf 17} (2010) 493-506, arXiv:1012.2635\\
A. Mironov, A. Morozov, A. Sleptsov, Theor.Math.Phys. {\bf 177} (2013) 1435-1470 (Teor.Mat.Fiz. {\bf 177} (2013)
179-221), arXiv:1303.1015; European Physical Journal, {\bf C73} (2013) 2492, arXiv:1304.7499;
Nuclear Physics, {\bf B889} (2014) 757-777, arXiv:1310.7622

\bibitem{AMMN}
A. Alexandrov, A. Mironov, A. Morozov, S. Natanzon
JHEP {\bf 11} (2014) 080, arXiv:1405.1395

\bibitem{IMM2}
H. Itoyama, A. Mironov, A. Morozov,
Nucl.Phys. {\bf B932} (2018) 52-118,
arXiv:1710.10027

\bibitem{MMNbg}
A.~Mironov, A.~Morozov and S.~Natanzon,
  Eur.\ Phys.\ J.\ {\bf C73} (2013) 2324,
arXiv:1208.5057\\
  A.~Mironov, A.~Morozov and S.~Natanzon,
  J.\ Geom.\ Phys.\  {\bf 73} (2013) 243,
arXiv:1210.6955\\
 A.~Mironov, A.~Morozov and S.~Natanzon, Journal of Knot Theory and Its Ramifications,
{\bf 23} (2014) 1450033

\bibitem{Schur} I. Schur,
J. Reine Angew. Math. {\bf 139} (1911) 155-250

\bibitem{Mac}  I.G. Macdonald,
{\it Symmetric functions and Hall polynomials}, Second Edition, Oxford University Press,
1995

\bibitem{Serg} A. Sergeev, 
Math. Sb. USSR, {\bf 51} (1985) 419–427

\bibitem{Sergrev} M. Yamaguchi, 
J. Algebra {\bf 222} (1999) 301–327, math/9811090\\
A. Kleshchev, {\sl Linear and projective representations of symmetric groups,} Cambridge
Tracts in Mathematics {\bf 163}, Cambridge Univ. Press (2005)

\bibitem{MMkerov}  S.V. Kerov, Func.An.and Apps. {\bf 25}  (1991) 78-81\\
A. Mironov, A. Morozov, arXiv:1811.01184

\bibitem{MM} A. Mironov, A. Morozov,
Phys.Lett. {\bf B771} (2017)  503-507, arXiv:1705.00976 \!\!;
Phys.Lett. {\bf B774} (2017) 210-216,  arXiv:1706.03667

\bibitem{MMchar} A. Mironov, A. Morozov,
JHEP {\bf 2018} (2018) 163,  arXiv:1807.02409

\bibitem{MMgenM} A. Mironov, A. Morozov, arXiv:1907.05410

\bibitem{Mormac}
A. Morozov, Phys.Lett. {\bf B785} (2018) 175-183,  arXiv:1808.01059 \!\!;
arXiv:1810.00395 \!\!;  arXiv:1901.02811

\bibitem{Hur} A. Hurwitz, 
Math. Ann. {\bf 38} (1891)
452–45

\bibitem{Fro}  G. Frobenius, 
Berl. Ber. {\bf 1896} (1896) 985–1021\\
W. Burnside, {\sl Theory of groups of finite order,} 2nd edition, Cambridge Univ. Press
(1911)\\
R. Dijkgraaf,
In: {\sl The moduli
spaces of curves},
Progress in Math. {\bf 129} (1995) 149-163,
Brikh\"auser

\bibitem{AN} A. Alexeevski, S. Natanzon, 
    Selecta Math., New ser.
{\bf 12:3} (2006) 307-377, math.GT/020216; Izv. Math. {\bf 72} (2008) 627-646, arXiv:0709.3601

\bibitem{Fulton} W. Fulton, {\sl Young tableaux: with applications to representation theory and geometry},
London Mathematical Society, 1997

\bibitem{EOP} A. Eskin, A. Okounkov, R. Pandharipande,
Adv. Math. {\bf 217} (2008) 873-888

\bibitem{G} S. Gunningham, 
Geom. Topol. {\bf 20} (2016) 1859-1907, arXiv:1201.1273

\bibitem{GD} D. Goulden , D.M. Jackson, A. Vainshtein,
Ann. of Comb. {\bf 4} (2000) 27-46,
Brikh\"auser, math/9902125

\bibitem{MMN1} A. Mironov, A. Morozov, S. Natanzon, Journal of Geometry and Physics, {\bf 62} (2012) 148-155, arXiv:1012.0433

\bibitem{IK} V. Ivanov, S. Kerov, 
Journal of Mathematical Sciences (Kluwer) {\bf 107} (2001) 4212-4230, math/0302203

\bibitem{OO} A.Yu. Okounkov, G.I. Olshanski, 
St. Petersburg Math. Journ. {\bf 9}
(1997) 73–146, q-alg/9605042

\bibitem{JM} E. Date, M. Jimbo, M. Kashiwara, T. Miwa, 
RIMS Symp. {\sl "Non-linear integrable
systems - classical theory and quantum theory"} (World Scientific,
Singapore, 1983)

\bibitem{GKLMM} A. Gerasimov, S. Khoroshkin, D. Lebedev, A. Mironov, A. Morozov,
Int.J.Mod.Phys. \textbf{A10} (1995) 2589-2614,
hep-th/9405011\\
A. Mironov, hep-th/9409190; Theor.Math.Phys. {\bf 114} (1998) 127, q-alg/9711006

\bibitem{Hirota}  R. Hirota, Phys.Rev.Lett. {\bf 27} (1971) 1192

\bibitem{GKM2} S. Kharchev, A. Marshakov, A. Mironov, A. Morozov, Int.J.Mod.Phys. {\bf A10} (1995) 2015, hep-th/9312210

\bibitem{Oko} A. Okounkov,
Math.Res.Lett. {\bf 7}
(2000) 447-453, math/0004128

\bibitem{OS} A. Orlov, D.M. Shcherbin,
Theor.Math.Phys.
{\bf 128} (2001) 906-926\\
A. Orlov,
Theor.Math.Phys. {\bf 146}
(2006) 183–206

\bibitem{AMMN1} A. Alexandrov, A. Mironov, A. Morozov, S. Natanzon,
J.Phys. A: Math.Theor. {\bf 45} (2012) 045209,
arXiv:1103.4100

\bibitem{MMMM} Yu. Makeenko, A. Marshakov, A. Mironov, A. Morozov, Nucl.Phys. {\bf B356} (1991) 574-628\\
A. Alexandrov, A. Mironov, A. Morozov,  JHEP {\bf 0912} (2009) 053, arXiv:0906.3305

\bibitem{GMMMO} A. Gerasimov, A. Marshakov, A .Mironov, A. Morozov, A. Orlov,
Nucl.Phys. {\bf B357} (1991) 565

\bibitem{KMMOZ} S. Kharchev, A. Marshakov, A. Mironov, A .Orlov, A. Zabrodin,
Nucl.Phys. {\bf B366} (1991) 569-601

\bibitem{Atiyah} M.F. Atiyah , 
Ann. Sci. Ecole Norm. Sup. {\bf 4}
(1971) 47–62

\bibitem{Mum} D. Mumford , 
Ann. Sci. École Norm. Sup. {\bf 4} (1971) 181–192

\bibitem{J} D. Johnson, 
J. London Math. Soc. (2) {\bf 22} (1980) 365-373

\bibitem{N2} S.M. Natanzon,  
Russian Math. Surveys, {\bf 54} (1999) 61-117

\bibitem{N3} S.M. Natanzon, {\sl Moduli of Riemann surfaces, real algebraic curves, and their superanalogs,} Translations of Mathematical Monographs, {\bf 225}, American Mathematical Society, Providence, RI, 2004, viii+160 pp

\bibitem{Lee}  J. Lee, T.H. Parker, 
Comm. Anal. Geom. {\bf 21} (2013) 1015 – 1060; arXiv:1212.1825\\
J. Lee, arXiv:1407.0055

\bibitem{Jose} T. J\'ozefiak, 
J. Pure Appl. Algebra {\bf 152} (2000) 187–193,

\bibitem{Mcauchy} A. Morozov,  arXiv:1812.03853

\bibitem{Ok} A. Okounkov, q-alg/9608021

\bibitem{BKP} E. Date, M. Jimbo, M. Kashiwara, T. Miwa,
Physica {\bf D4} (1982) 343-365\\
M. Jimbo, T. Miwa,
Publ. RIMS Kyoto Univ. {\bf 19} (1983) 943-1001

\bibitem{BKPs} Y. You,
Adv. Ser. Math. Phys. {\bf 7} (1990) 449- 464, World Science Publishing, Teaneck, New Jersey\\
J.J.C. Nimmo,
J. Phys. {\bf A23} (1990) 751-760

\bibitem{Orlov} A. Orlov,  Theor. Math. Phys. {\bf 137} (2003) 1574-1589, math-ph/0302011\\
J.J.C. Nimmo, A. Orlov, Glasgow Mathematical Journal {\bf 47 (A)} (2005) 149-168, nlin/0405009\\
J. Harnad, J.W. van de Leur, A.Yu. Orlov, Theor. Math. Phys. {\bf 168} (2011) 951-962, arXiv:1101.4216\\
J.W. van de Leur, A.Yu. Orlov, arXiv:1404.6076; arXiv:1611.04577\\
A.Yu. Orlov, T. Shiota, K. Takasaki, arXiv:1201.4518; arXiv:1611.02244

\bibitem{IMM} H. Itoyama, A. Mironov, A. Morozov, Phys.Lett. {\bf B788} (2019) 76-81, arXiv:1808.07783

\end{thebibliography}
\end{document}